\documentclass{aa}
\usepackage{graphicx,subfigure,natbib,url,txfonts,epsfig}
\bibpunct{(}{)}{;}{a}{}{,}

\begin{document}
\title{Reconstruction of total solar irradiance 1974 - 2009}

\author{William T. Ball\inst{1}, Yvonne C. Unruh\inst{1}, Natalie A. Krivova\inst{2}, Sami Solanki\inst{2,3}, Thomas Wenzler\inst{4}, Daniel J. Mortlock\inst{1,5}, Andrew H. Jaffe\inst{1}}
        \offprints{william.ball08@imperial.ac.uk}
	\institute{Astrophysics Group, Blackett Laboratory, Imperial College London, London, SW7 2AZ, United Kingdom
   \and 
	Max-Planck-Institut f\"ur Sonnensystemforschung, D-37191 Katlenburg-Lindau, Germany
	\and
	School of Space Research, Kyung Hee University, Yongin, Gyeonggi, 446-701, Korea
	\and
	ZHAW Zurich University of Applied Science, CH-8401 Winterthur, Switzerland
	\and
	Department of Mathematics, Imperial College London, London, SW7 2AZ, United Kingdom
              }

\abstract
{The study of variations in total solar irradiance (TSI) is important for understanding how the Sun affects the Earth's climate.}
{Full-disk continuum images and magnetograms are now available for three full solar cycles. We investigate how modelled TSI compares with direct observations by building a consistent modelled TSI dataset. The model, based only on changes in the photospheric magnetic flux can then be tested on rotational, cyclical and secular timescales.} 
{We use Kitt Peak and SoHO/MDI continuum images and magnetograms in the SATIRE-S model to reconstruct TSI over cycles 21-23. To maximise independence from TSI composites, SORCE/TIM TSI data are used to fix the one free parameter of the model. We compare and combine the separate data sources for the model to estimate an uncertainty on the reconstruction and prevent any additional free parameters entering the model.}
{The reconstruction supports the PMOD composite as being the best historical record of TSI observations, although on timescales of the solar rotation the IRMB composite provides somewhat better agreement. Further to this, the model is able to account for 92\% of TSI variations from 1978 to 2009 in the PMOD composite and over 96\% during cycle 23. The reconstruction also displays an inter-cycle, secular decline of 0.20$^{+0.12}_{-0.09}$ Wm$^{-2}$ between cycle 23 minima, in agreement with the PMOD composite.}
{SATIRE-S is able to recreate TSI observations on all timescales of a day and longer over 31 years from 1978. This is strong evidence that changes in photospheric magnetic flux alone are responsible for almost all solar irradiance variations over the last three solar cycles.}


\titlerunning{TSI Reconstruction, 1974-2009}
\authorrunning{Ball et al.}
\maketitle

\section{Introduction} 
%
\label{sec:intro}

The Sun is the primary source of radiation entering the Earth's climate system, reaching to all levels of the atmosphere and into the ocean \citep{Haigh1994, BondKromer2001, LabitzkeAustin2001}. While much of the modulation from the Sun seen in the Earth's atmosphere is a result of variation in spectral solar irradiance (SSI), total solar irradiance (TSI) encompasses the combined variation across all wavelengths and thus is a measure of the total forcing. Consequently, TSI is an important quantity in climate studies.

TSI is also easier to measure than SSI and should, in principle, provide a more stable and accurate record of solar variability. However, while direct, above-atmosphere, radiometric observations exist from 1978, continuous observations have required the use of several satellites. This has lead to an inevitable need to account for different absolute level calibrations, degradation corrections and switch-off periods in each instrument. Accounting for each of these problems naturally leads to some interpretation and, as a result, three competing TSI composites have been proposed \citep{FrohlichLean1998, WillsonMordvinov2003, DewitteCrommelynck2004}.

\cite{WenzlerSolanki2009} and \cite{KrivovaSolanki2009c} have shown that the PMOD composite \citep{FrohlichLean1998} is most likely the best record of TSI, but this required fitting a model to each composite separately and was not independent of the composites. Also, the reconstructions used in these investigations ended in 2003 and hence do not cover the unusual minimum between cycles 23 and 24.

Many studies have shown that the Sun's surface magnetic flux, in the form of sunspots and faculae, can explain TSI variation on \emph{rotational} time scales of days and weeks \citep{WillsonGulkis1981,FoukalLean1986,FliggeSolanki2000} and \emph{cyclical} periods covering the solar cycle \citep{LeanCook1998,WenzlerSolanki2006}. All three TSI composites suggest that the Sun undergoes some measure of inter-cycle, \emph{secular} variability, but all three are in disagreement as to the magnitude and direction of this change. Some indices, e.g. open solar flux \citep{LockwoodBell2010}, show that the minimum of 2008 saw the largest significant secular change of the satellite era and this is reflected in the PMOD composite by a decline in TSI relative to the minima of 1996 and 1986. Although the decline in TSI between 1996 and 2008 is small, estimated to be $\sim$0.2 Wm$^{-2}$ \citep{Frohlich2009b}, the importance of this small change lies in the accumulated change over centuries. It has been suggested that the Sun may be entering a grand minimum state, though there is still some debate on this \citep{FeymanRuzmaikin2011,SolankiKrivova2011}.

To understand the impact of the Sun on climate change, it is imperative that the physical mechanism behind the variation of the Sun on secular timescales is understood to a sufficient level to be able to confidently reconstruct solar variability into the past. This would then allow reconstructions that extrapolate TSI proxy data, such as sunspot number and the geomagnetic aa index, to be refined and improved \citep{TappingBoteler2007,SteinhilberBeer2009,VieiraSolanki2010}. The potential range for a secular change in TSI over the last four hundred years, since the Maunder Minimum, could be significant in climate forcing and a better understanding of the Sun during the satellite era could lead to more precise estimates of the Sun's true effect on the Earth over that period, although there are diverging views on this. On the one hand, \cite{ShapiroSchmutz2011} argue that because the Sun has been in a particularly active phase during the last century, one cannot currently extrapolate the full range of solar variability on secular scales. On the other hand, \cite{SchrijverLivingston2011} suggest that solar activity during the 2008 minimum has reached levels similar to the Maunder Minimum, a period during which sunspots were conspicuous by their absence, and cannot go much lower.

Explanations for secular trends other than solar surface magnetic activity have also been proposed. E.g., \cite{RozelotLefebvre2004} suggested that changes in the Sun's radius could account for long-term changes in the Sun. \cite{HarderFontenla2009} proposed that there may have been a change in the photospheric temperature gradient that produced a spectrally dependent change, increasing irradiance at some wavelengths and decreasing at others. A global temperature change has also been proposed \citep{TappingBoteler2007} with \cite{Frohlich2009b} concluding that the recent minimum was caused by a global temperature decline of 0.2 K in the Sun's effective temperature.

The SATIRE-S model, denoted with an `S' for the satellite era (see \cite{KrivovaSolanki2011} to distinguish between the various SATIRE types), is a well established model that has proven very successful. Using Kitt Peak full-disk data, between 1974 and 2003, the model has been able to reproduce 83\% of TSI variability \citep{WenzlerSolanki2006}. Using SoHO data, \cite{KrivovaSolanki2003} and \cite{BallUnruh2011} were able to account for 92\% and 97\% during the ascending and descending phases of cycle 23, respectively.

\cite{Steinhilber2010} used recently recalibrated MDI synoptic data within a model framework similar to that of the SATIRE model over the full period of cycle 23, but was unable to reproduce the recent decline seen by the PMOD composite. The cutoff threshold for low magnetic flux used was higher than previous reconstructions and could have removed a significant portion of the relatively weak magnetic flux in the network that might be responsible for secular variation.

In this study, we combine all available full disk continuum images and magnetograms from Kitt Peak (KP) and the Solar and Heliospheric Observatory (SoHO) to compute TSI for a period of 36 years, extending the reconstruction, from 1974 to 2009.

In section~\ref{sec:model} we briefly outline the SATIRE-S model. In section~\ref{sec:data} we describe the full-disk images used in the model, the TSI data used to set up the reconstruction and the TSI composite data used to make comparisons with. Continuum images and magnetograms from different sources each have their own intricacies that need to be accounted for before they can be combined. We explain how this is done in section~\ref{sec:homog}. The datasets are combined to produce a consistent three-cycle reconstruction in section~\ref{sec:recon}. Comparisons are made between the reconstruction and TSI composites in section~\ref{recon:compare} and discussion of the results and the conclusions are presented in section~\ref{sec:discuss}.

\section{Modelling Solar Irradiance using SATIRE-S}
\label{sec:model}
Solar irradiance is modelled using the SATIRE-S model that assumes that only changes in surface magnetic flux are responsible for changes in solar irradiance on time scales greater than one day. \cite{FliggeSolanki2000}, \cite{KrivovaSolanki2003} and \cite{WenzlerSolanki2004,WenzlerSolanki2005,WenzlerSolanki2006} explain the model in depth and \cite{BallUnruh2011} cover recent modifications. A detailed description can be found in these papers and references therein; here we give a short description only.

SATIRE-S is a four-component model that uses daily continuum intensity images and magnetograms. After correcting for limb darkening, sunspot penumbra and umbra are identified by contrast level in continuum intensity images. Faculae are identified in magnetograms, where no spot pixels are located, above a threshold determined by the noise level, B$_{\mathrm{thr}}$ = 3$\sigma_{\mathrm{noise}}$ (see sections~\ref{data:satire} and \ref{sec:mdimag}). Note that a line-of-sight correction is applied to the magnetograms, assuming that observed flux is radial in nature. This can produce artifacts at the limb, so all pixels with a limb angle of $\mu = \cos\theta$ $< 0.1$, where $\theta$ is the heliocentric angle, are ignored. All remaining pixels are considered to be the quiet Sun. 

Faculae are composed of magnetic elements too small to be resolved in the full disk magnetograms, so it is assumed that the facular filling factor increases linearly with increasing magnetic flux up to a saturation point, B$_{\mathrm{sat}}$, above which the pixel is assumed to be `full' and the filling factor is held at unity (cf. \citeauthor{FliggeSolanki2000}, \citeyear{FliggeSolanki2000}). B$_{\mathrm{sat}}$ is the only freely varying parameter in the model and is used to achieve good agreement with TSI observations. Magnetogram pixels of high magnetic flux are incorrectly assigned as faculae, so a cutoff, B$_{\mathrm{cut}}$, was introduced in \cite{BallUnruh2011}. This fixed parameter is discussed in section~\ref{homog:darkpix}.

Intensities for each component, depending on $\mu$ and on the wavelength $\lambda$, are determined and a model atmosphere is employed for each respective component as set out in \cite{UnruhSolanki1999}. The resulting spectrum is integrated on each day to produce relative variations in TSI.

\section{Data}
\label{sec:data}

\subsection{Data used in SATIRE-S}
\label{data:satire}
We use full-disk continuum intensity images and magnetograms employed in SATIRE-S, taken from two locations: the National Solar Observatory Kitt Peak Vacuum Tower (KP) and the Solar and Heliospheric Observatory (SoHO). KP data were taken using the 512-channel diode array (KP/512) from 1974. KP/512 was replaced by the spectromagnetograph (KP/SPM) which operated between 1993 and 2003. The Michelson Doppler Imager (MDI) operated onboard the SoHO satellite between 1996 and 2011. 

\subsubsection{SoHO/MDI}
\label{data:mdi}
MDI \citep{ScherrerBogart1995} records 1024 $\times$ 1024 pixel full-disk filtergrams and magnetograms in the Ni\,I absorption line at 676.78 nm with 4'' resolution. Simultaneous continuum intensity image and magnetogram data are usually not available to download. Therefore, continuum intensity images taken within 12 hours of the magnetograms are chosen to avoid the introduction of large differences resulting from surface evolution or from rotation off the visible disk. The continuum intensity images are then rotated to align features in both images. Continuum intensity images have a measurement uncertainty of 0.3\% and only a single image is needed to identify the sunspot components. For the magnetograms, we attain a reduced noise level by using 5-minute averaged magnetograms. The calibrated 1-minute magnetograms have a reported uncertainty of $\sim$30 G; averaging over five-minutes reduces the uncertainty to $\sim$15 G (these are recalibrated values following \citep{TranBertello2005}; see below).

In this study, images are used from almost the entire working period of the MDI instrument from 1996 May 20 to 2009 October 31, using 4140 image pairs.

Previous studies using SATIRE-S have employed level 1.5 magnetograms \citep{FliggeSolanki2000,KrivovaSolanki2003,WenzlerSolanki2004}. Recalibrations of the data have now produced the level 1.8.2 magnetogram data used in \cite{BallUnruh2011}, \cite{KrivovaSolanki2011b} and this work. The most significant change came after \cite{TranBertello2005} compared the sensitivity of MDI magnetograms with those of the Mount Wilson Observatory. This work led to a position-dependent correction\footnote{All modifications to produce level 1.8.2 magnetograms are described in more detail at http://soi.stanford.edu/magnetic/Lev1.8/} that has increased the estimated observed magnetic flux by a factor of $\sim$1.6, necessitating a change in SATIRE-S parameters, notably B$_{\mathrm{sat}}$.

SoHO was subject to a number of operational problems that MDI was not exempt from. It appears that these, along with some operational changes to the instrument, have resulted in changes to the magnetograms over time. We discuss the causes and effects of these changes in section~\ref{sec:mdimag}.

\subsubsection{KP/SPM}
\label{data:spm}
KP/SPM \citep{JonesHarvey1992} operated between 1992 and 2003 taking 1788 $\times$ 1788 pixel images in the Fe I line at 868.8 nm with a 1.14'' pixel size. Continuum intensity images and magnetograms were extracted from slit spectra that scanned across the solar disk in one hour. \cite{JonesHarvey1992} and \cite{WenzlerSolanki2004} both determined a background noise level of $\sim$5 G $\simeq$ $\sigma_{\mathrm{noise,SPM}}$. Note that terms with the subscripts MDI, SPM and 512 mean they are set for reconstructions using only that specific dataset.

\cite{WenzlerSolanki2005} identified 2055 sets of images between 1992 November 21 and 2003 September 21 that were relatively free of atmospheric effects. We use these images in this study. 

\subsubsection{KP/512}
\label{data:512}
Observations by the KP/512 magnetograph \citep{LivingstonHarvey1976} were taken between 1974 and 1993. Full-disk images are 2048 $\times$ 2048 pixels with a 1'' pixel size, but with a generally siginificantly lower spatial resolution given by seeing. Continuum intensity images and magnetograms were built simultaneously from scans lasting approximately 40 minutes. Continuum intensity images are limited to a 4-bit intensity scale providing only 15 discrete levels that make the separation of penumbra from umbra, or sunspots from limb-darkening, difficult. A brief description of the procedure employed by \cite{WenzlerSolanki2006} to resolve this is given in section~\ref{homog:spot}. \cite{JonesHarvey1992} and \cite{WenzlerSolanki2006} both found that the magnetograms had a background noise level of $\sim$7-9 G; \cite{WenzlerSolanki2006} found a mean of 8.1 G = $\sigma_{\mathrm{noise,512}}$, which we adopt here.

As the observatory is ground-based, seeing and weather conditions in general have a significant effect on the quality of the images. Work done by \cite{WenzlerSolanki2006} identified 1734 dates with usable data over the period 1974 February 1 to 1992 April 4. In 1992 there was a change of instrument and KP/512 observations are only available as an overlap with the superceding instrument, KP/SPM, on 45 days between 1992 November 28 and 1993 April 10.

\cite{WenzlerSolanki2006} found that a conversion factor, f$_{\mathrm{512}}$, was necessary in order to multiply magnetic flux in KP/512 magnetograms to bring them into agreement with KP/SPM. The official value of f$_{\mathrm{512}}$ quoted in the data release was 1.46, but \cite{WenzlerSolanki2006} found that to get best agreement with PMOD a value of 1.60 was needed. The conversion factor allows other parameters, such as the magnetic flux level B$_{\mathrm{thr}}$ and B$_{\mathrm{sat}}$, to be the same in both KP/SPM and KP/512 reconstructions. We show, in section~\ref{homog:f512}, that f$_{\mathrm{512}}$ is dependent on B$_{\mathrm{sat}}$ and has no freedom to vary independently.

\subsection{Observational TSI composite}
\label{data:pmod}
TSI observations are needed to fix the free parameter, B$_{\mathrm{sat}}$, in SATIRE-S. They also provide a basis with which to test the model reconstruction on various timescales.

TSI observations began in 1978, but no single satellite has observed the whole period. Therefore, datasets have been combined into composites. There are currently three published composites adopting different approaches to combine the radiometric data and account for degradation, switch-offs and sudden glitches. These composites are known as PMOD \citep{Frohlich2003}, ACRIM \citep{WillsonMordvinov2003} and IRMB \citep{DewitteCrommelynck2004}. In section~\ref{recon:compare} we compare the three-cycle TSI reconstruction with these composites.

The current version of the PMOD composite, d41\_62\_1003 \citep{Frohlich2000}, covers the period 1978 to present. It is composed of data from four radiometers: NIMBUS7/HF, SMM/ACRIM I, UARS/ACRIM II and the PMO6V radiometer on the SoHO/VIRGO instrument (see references within \citeauthor{FrohlichLean1998}, \citeyear{FrohlichLean1998}). The ACRIM composite also uses NIMBUS7/HF and SMM/ACRIM I, but where PMOD uses SoHO/VIRGO from 1996, the ACRIM composite continues to use UARS/ACRIM II until 2000 before switching to ACRIMSAT/ACRIM III. The latest version of the IRMB composite begins in 1984, employing NIMBUS7/HF, SMM/ACRIM I, UARS/ACRIM II and then the DIARAD radiometer, on the SoHO/VIRGO instrument, from 1996.

The current absolute irradiance level of PMOD is set at $\sim$1365 Wm$^{-2}$ during the solar minimum of 1986 \citep{CrommelynckFichot1995}. The latest TSI instrument, SORCE/TIM, has observed irradiance at $\sim$4.5 Wm$^{-2}$ lower than VIRGO; recent calibration measurements have confirmed this value to be more accurate \citep{KoppLean2011}. However, the relative variation of the PMOD composite is in good agreement with SORCE/TIM and it is the relative variation that is our main concern in this study; SATIRE-S can be scaled to either level. For these reasons, we normalise all three composites and SATIRE-S to the absolute level of TIM at solar minimum in 2008.

\subsection{SORCE/TIM}
\label{data:tim}
The Total Irradiance Monitor (TIM), a radiometer on the SORCE satellite, has been observing TSI since February 2003 with an absolute accuracy of 350 ppm, an instrumental noise of $\sim$2 ppm and a stability of better than 10 ppm/yr \citep{KoppLawrence2005, KoppLawrence2005b, KoppLean2011}. Apart from the lower absolute value of TSI, TIM also displays a shallower gradient than PMOD during the declining phase of cycle 23. In this study we use version 11, level 3 six-hourly data\footnote{SORCE/TIM data downloaded 2011 July 11} averaged onto a daily cadence.

By the time solar minimum had been reached, in December 2008, TIM had been operating for about 5.5 years. This equates to a long-term 1$\sigma$ uncertainty in the change in TSI of $\sim$55 ppm or $\sim$0.07 Wm$^{-2}$. This is about the same as the difference in TSI variations seen between VIRGO and TIM over the same period and means that the gradients in TSI agree with each other within this uncertainty, not even taking into account the error on VIRGO (private communication, Greg Kopp). 

\section{Homogenising KP/512, KP/SPM and MDI data}
\label{sec:homog}

\begin{figure*}[!ht]
\centering
 \resizebox{\hsize}{!}{\includegraphics[width=1.\textwidth, angle=0]{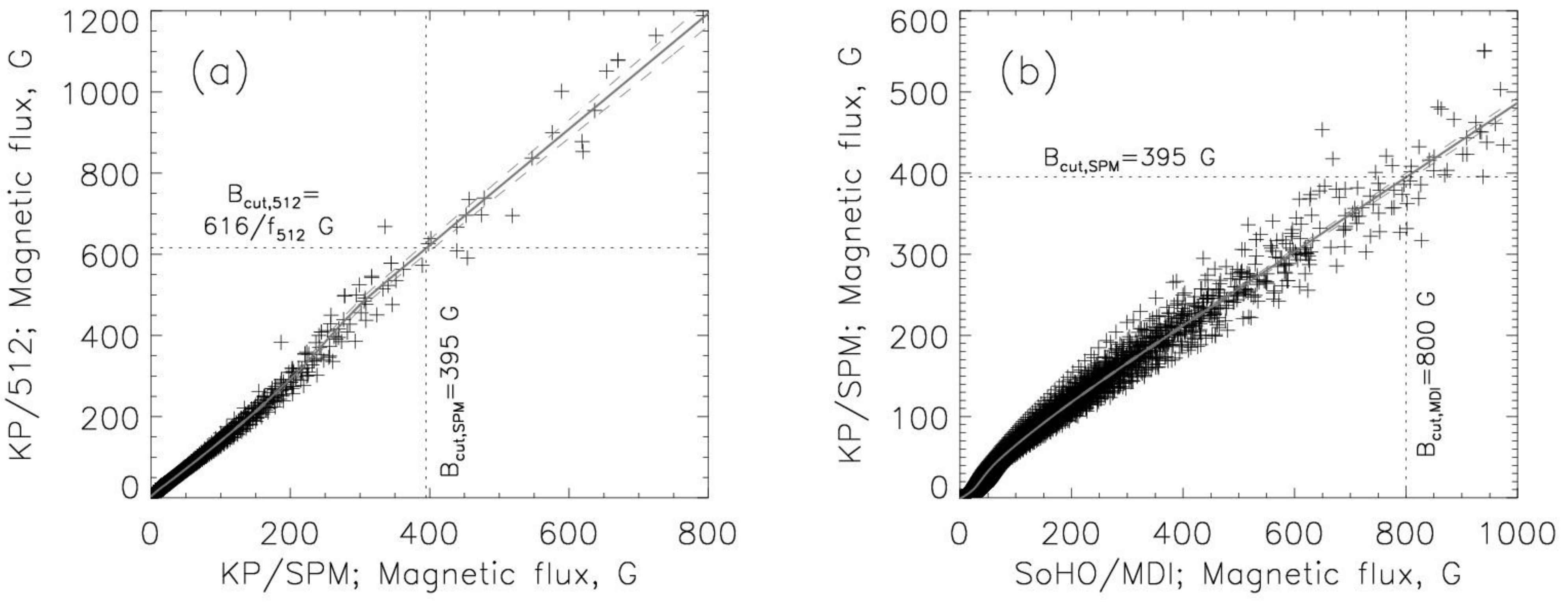}}
 \caption[]{Histogram equating plots. (\textbf{a}) 18 contemporaneous KP/SPM and KP/512 magnetograms and (\textbf{b}) 84 contemporaneous SoHO/MDI and KP/SPM magnetograms. The grey line is the interpolated mean of the black crosses with the dashed grey lines the standard error. Dotted lines highlight the relationship of B$_{\mathrm{cut}}$ between the datasets.}
 \label{fig:histeq1}
\end{figure*}

In this study, we combine KP/512 and KP/SPM with SoHO/MDI data. The studies of \cite{KrivovaSolanki2003} and \cite{WenzlerSolanki2004,WenzlerSolanki2005,WenzlerSolanki2006} also used these data, but since these publications, MDI data have been recalibrated (see section~\ref{data:mdi}). This means that settings in SATIRE-S used by these authors may no longer be appropriate and need to be reconsidered. 

In this section, we discuss how the parameters in SATIRE-S are determined so that consistency is achieved across all three datasets.

\subsection{Sunspot areas and temperatures}
\label{homog:spot}
The study of \cite{WenzlerSolanki2006} combined KP/SPM and KP/512 data. They set sunspot thresholds so that agreement was found, in both datasets, with an independent ground-based composite record of sunspot area \citep{BalmacedaSolanki2005}. The KP/SPM penumbral threshold was set in \cite{WenzlerSolanki2004} at 0.92 so as to obtain good agreement with total sunspot area. \cite{Wenzler2005PHD} confirmed that the ratio of umbral area to total sunspot area is 0.2 and used this to determine a KP/SPM umbral threshold of 0.70. 

In this study, we use the KP/512 sunspot data computed by \cite{WenzlerSolanki2006} for the KP/512 part of the reconstruction. The way these were obtained by \cite{WenzlerSolanki2006} is as follows. Due to the 4-bit digitization of the KP/512 continuum intensity images, only the combined total umbral and penumbral sunspot areas could be identified. Therefore, \cite{WenzlerSolanki2006} used the established ratio of umbra to sunspot area of 0.2 to estimate umbral areas. Pores could not be identified, so that a simple multiplication of the total sunspot area brought the KP/512 sunspot areas into agreement with an independent sunspot area composite.

Following from this, we vary MDI contrast thresholds to bring them into agreement with the spot ratio of 0.2 and KP/SPM. Using 847 continuum images from KP/SPM and MDI between 1999 and 2003, we compare areas and find that the penumbral and umbral thresholds should be 0.89 and 0.64 respectively. The difference between KP and MDI thresholds is partly due to the different wavelengths at which the instruments operated, as well as differences in stray light.

The total effect of sunspots on TSI is degenerate between the temperature, T$^{4}$, of the model atmosphere and the area of the identified sunspot. For consistency we use temperatures of 5450 K for penumbra and 4500 K for umbra throughout, as in \cite{WenzlerSolanki2006}.

\subsection{The KP/512 magnetogram conversion factor, f$_{\mathrm{512}}$}
\label{homog:f512}

\begin{figure*}[!ht]
\centering
 \resizebox{\hsize}{!}{\includegraphics[angle=180,bb = 75 275 715 545]{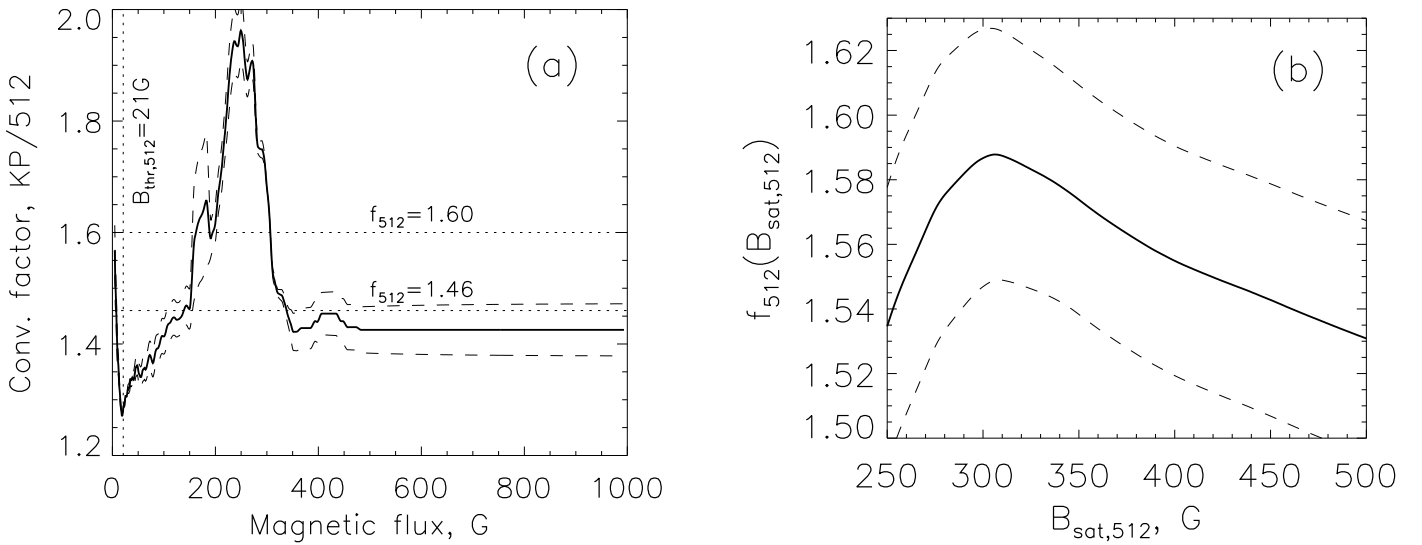}}
 \caption[]{(\textbf{a}) The factor by which magnetic flux in KP/512 magnetograms should be multiplied in order to obtain the same magnetic flux as in KP/SPM magnetograms, as a function of magnetic flux (see text). (\textbf{b}) The best global conversion factor, f$_{\mathrm{512}}$, by which complete KP/512 magnetograms should be multiplied as a function of B$_{\mathrm{sat,512}}$, the SATIRE-S free paramter. Dashed lines in both panels represent the standard error as shown in Fig.~\ref{fig:histeq1}a.}
 \label{fig:histeq2}
\end{figure*}

The introducion of f$_{\mathrm{512}}$ (see section~\ref{data:512}) by \cite{WenzlerSolanki2006} to easily combine KP/512 and KP/SPM created an additional free parameter in the reconstruction. We are able to remove the freedom in this parameter by considering how f$_{\mathrm{512}}$ relates to B$_{\mathrm{sat}}$. This is done as follows.

Magnetic flux is registered differently in KP/512 and KP/SPM magnetograms, so we use the method of histogram equating \citep{JonesCeja2001,WenzlerSolanki2004} to compare the signal observed in the magnetograms. In Fig.~\ref{fig:histeq1}a, we compare unsigned magnetic flux in 18 KP/SPM and KP/512 magnetograms taken on the same days in 1992/3. Crosses are the comparative magnetic flux calculated by sorting pixel values into ascending order and taking the mean in each of 1000 equally divided bins. By interpolating the KP/512 values onto a linear KP/SPM scale and averaging, the solid grey line results. This gives the corresponding mean level of magnetic flux in KP/SPM as observed by KP/512; the grey dashed lines show the standard error on the mean. Then, the mean level of magnetic flux over 10 G intervals centred on each level of magnetic flux along the grey line is calculated. The ratio of these mean magnetic flux values in KP/512 to those of KP/SPM result in a conversion factor at every level of magnetic flux, as shown in Fig.~\ref{fig:histeq2}a. In this plot, the solid line is the factor required to convert magnetic flux in KP/512 to bring it in to agreement with KP/SPM, as a function of magnetic flux. The dashed lines show the standard errors, also from Fig.~\ref{fig:histeq1}a. 

The official conversion factor value of f$_{\mathrm{512}}$ = 1.46 and the factor value of f$_{\mathrm{512}}$ = 1.60 found by \cite{WenzlerSolanki2006} are indicated by horizontal dashed lines in Fig.~\ref{fig:histeq2}a. It now becomes clear where the official conversion factor and that of \cite{WenzlerSolanki2006} come from: at high flux levels above 350 G the conversion factor is close to the official factor; below this it varies between 1.25 and 1.90, a range that very nearly averages to the value of \cite{WenzlerSolanki2006}. 

In SATIRE-S, magnetic flux below B$_{\mathrm{thr}}$ is ignored and all magnetic flux above B$_{\mathrm{sat}}$ is considered to have the same effect on irradiance, so only fluxes between these two values are important in determining f$_{\mathrm{512}}$. An approximate value for f$_{\mathrm{512}}$ can now be calculated by simply taking the mean of the conversion factor between B$_{\mathrm{thr,512}}$ and B$_{\mathrm{sat,512}}$. B$_{\mathrm{thr,512}}$ is fixed, therefore f$_{\mathrm{512}}$ depends only on the chosen value of B$_{\mathrm{sat,512}}$. The result is that f$_{\mathrm{512}}$ is completely determined by B$_{\mathrm{sat,512}}$, has ceased to be an independent free parameter, and allows us to combine KP/SPM and KP/512 images while maintaining the same value of the single free parameter, B$_{\mathrm{sat}}$.

Although B$_{\mathrm{thr,512}}$ = 21 G was determined using f$_{\mathrm{512}}$ = 1.46, it changes very little with f$_{\mathrm{512}}$, so we leave it at this value. In Fig.~\ref{fig:histeq2}b, the resulting conversion factor as a function of B$_{\mathrm{sat,512}}$ can be seen to lie in the range of B$_{\mathrm{sat,512}}$ of 250 to 500 G (solid line) with the propagated standard errors (dashed lines) from the left plot at approximately $\pm$ 0.04. In section~\ref{recon:combine} we find that the required value of B$_{\mathrm{sat,512}}$ is 330 G. The resulting value of f$_{\mathrm{512}}$, 1.580, is close to the best value determined by \cite{WenzlerSolanki2006} when the model was fitted to the PMOD composite. Now, however, f$_{\mathrm{512}}$ is determined independently of the PMOD composite.

\subsection{Removing dark high-flux magnetogram pixels}
\label{homog:darkpix}

\cite{BallUnruh2011} found that high magnetic flux facular pixels identified in MDI magnetograms were incorrectly enhancing reconstructed irradiance relative to TSI observations. By ignoring pixels at and above this level, B$_{\mathrm{cut,MDI}}$ = 800 G, the agreement between modelled irradiance and SORCE/TIM improves, especially on rotational timescales. \cite{BallUnruh2011} incorrectly assumed that the physical reason was due to a decrease in continuum contrast with increasing magnetic flux as in \cite{OrtizSolanki2002}. According to Kok Leng Yeo (private communication), the magnetic flux of pore and sunspot canopies are being detected near the limb and falsely attributed as faculae. The improving effect of removing these pixels appears to be most significant at $\sim$550 G, but down to this level real magnetic features away from the limb are also removed. Therefore, the more conservative cutoff level of 800 G is maintained in this paper.

For consistency, this result needs to be propagated through to KP/SPM and KP/512 magnetograms. Figure~\ref{fig:histeq1}b is constructed in the same way as Fig.~\ref{fig:histeq1}a, as described in the previous subsection. To calculate the shift, B$_{\mathrm{cut,SPM}}$, we compare 84 magnetograms observed within two hours of corresponding MDI magnetograms between 1999 and 2003. Using such close temporal proximity reduces the effect of evolving active regions. This indicates that when B$_{\mathrm{cut,MDI}}$ = 800 G then B$_{\mathrm{cut,SPM}}$ = 395 G. We repeat the same process for KP/512 using Fig.~\ref{fig:histeq1}a. It follows that B$_{\mathrm{cut,512}}$ = 616/f$_{\mathrm{512}}$  G, where f$_{\mathrm{512}}$ is the conversion factor (see section~\ref{data:512}).

We have established that B$_{\mathrm{cut}}$ is dependent on the chosen value in MDI. While this was a somewhat arbitrary choice, it is not a freely varying parameter. Since the freedom of f$_{\mathrm{512}}$ can be removed and made dependent on B$_{\mathrm{sat}}$ (see section~\ref{homog:f512}), no additional free parameters have been introduced here.

\subsection{Correcting pre-1990 KP/512 magnetograms}
\label{homog:shiftcorr}

\begin{figure*}[!ht]
\centering
 \resizebox{\hsize}{!}{\includegraphics[bb=40 25 600 300, angle=0]{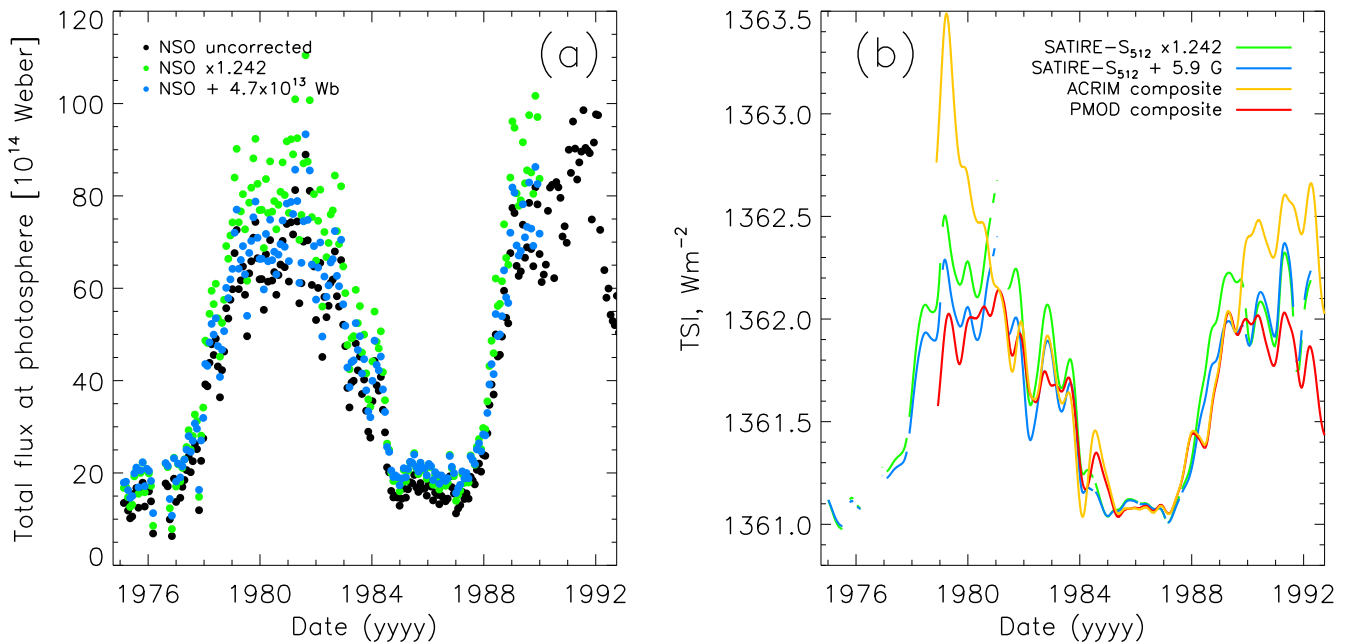}}
 \caption[]{(\textbf{a}) Total photospheric flux, obtained from KP/512 Carrington maps, before applying a correction (black dots), after either a scaling (green) or a shift is applied (blue). (\textbf{b}) SATIRE-S reconstruction of TSI using KP/512 continuum intensity images and magnetograms only, with either a scaling or a shift compared to the ACRIM and PMOD composites. All datasets are smoothed and normalised to the minimum of 1986. Gaps in the curves are present when data gaps are larger than 27 days. See main text for scaling and shifts applied to KP/512 Carrington maps and magnetograms.}
 \label{fig:shift_multi_plots}
\end{figure*}

\cite{ArgeHildner2002} reported that, prior to 1990, the total photospheric magnetic flux from KP/512 Carrington maps (NSO) was lower compared to that from Wilcox (WSO) and Mount Wilson (MWO) Solar Observatories Carrington maps. They used their analysis to develop a correction procedure, but this was only applied to a limited number of magnetograms, most of which were recorded during inactive times of the solar cycle. \cite{WenzlerSolanki2006} needed to use all available KP/512 magnetograms. They found a straight-line fit between the corrected and uncorrected total Carrington-rotation integrated photospheric flux, (see Fig. 3 of \citeauthor{WenzlerSolanki2006}, \citeyear{WenzlerSolanki2006}). The linear fit was excellent (correlation coefficient, r$_{\mathrm{c}}$=0.98), so all pre-1990 magnetograms were scaled by the regression slope value of 1.242 (the intercept of the fit was 0.306).

In Fig.~\ref{fig:shift_multi_plots}a we show the effect of scaling by a factor of 1.242 (green dots) compared to uncorrected Carrington maps (black dots), shown as the total photospheric flux of each Carrington roration between 1974 and 1992. The uncorrected Carrington maps can be seen compared to the WSO and MWO in Fig. 1 of \cite{ArgeHildner2002}. Figure 5 of \cite{WenzlerSolanki2006} shows the effect after scaling KP magnetograms by a factor of 1.242 and while the scaling seems to have improved the NSO results during the 1974-1976 minimum relative to WSO and MWO, it appears that the effect is too large between 1987 and 1989. The effect on the maximum of cycle 21 seems to be reasonable but in general the scaled level sits above MWO and WSO level of total flux at all other times. This may be due to the fact that the scaling factor is mainly derived from Carrington maps recorded during solar activity minima. Extrapolating the scaling factor to more active periods may cause magnetic flux at cycle maximum to be overestimated. Therefore, we also consider an alternate correction to apply.

While a general correction to pre-1990 data would likely take the form of a time-dependent shift and multiplication, this would require a detailed analysis of all magnetograms. In this paper, we consider the effect of applying both a constant shift or a scaling. To calculate, we take the difference between uncorrected and corrected total NSO flux below 3 $\times 10^{15}$ Weber during low activity times. The value of 3 $\times 10^{15}$ Weber is arbitrary but encompasses the majority of low activity Carrington maps. The mean difference is 4.7 G $\times 10^{14}$ Weber. The effect of applying this shift to the Carrington maps is also shown in Fig.~\ref{fig:shift_multi_plots}a as blue dots. At solar minima, the shift results in similar values to the scaling, but the cycle amplitude is lower and agrees better with MWO and WSO during the early part of cycle 22 maximum. Unfortunately, the shift derived for the synoptic charts cannot be converted to a shift in the KP/512 magnetograms: total magnetic flux reported in KP/512 magnetograms can exceed NSO Carrington maps by a factor of well over 100 due to massive flux cancellation in the lower-resolution Carrington maps.

Instead, we take a different approach to calculate a constant shift for the pre-1990 KP/512 magnetograms. First, we assume that the scaling factor of 1.242 found by \cite{WenzlerSolanki2006} is appropriate for low activity magnetograms. We can use the difference between low-level magnetic flux in uncorrected and scaled magnetograms as a constant shift. For this we use the threshold level determined by \cite{WenzlerSolanki2006} for the KP/512 magnetograms, i.e. B$_{\mathrm{thr,512}}$ = 3$\sigma_{mag,512}$ = 3 $\times$ 8.1 G = 24.3 G. This value was found by using only magnetograms during the solar minimum of 1986 and, therefore, is appropriate for use in calculating a constant shift. The difference between scaled and unscaled B$_{\mathrm{thr,512}}$ is 5.9 G. In Fig.~\ref{fig:shift_multi_plots}b, reconstructions using KP/512 magnetograms only are shown, with scaled in green and shifted in blue. All datasets in this plot are shown as smoothed time series and normalised to the solar minimum of 1986. All smoothings of plotted time series throughout the whole paper, and for detrending data, are performed with a Gaussian window with an equivalent boxcar width of 135 days. The result of adding a constant instead of scaling is to reduce the cycle amplitude prior to 1990. 

\begin{figure*}[!ht]
\centering
 \resizebox{\hsize}{!}{\includegraphics[angle=0]{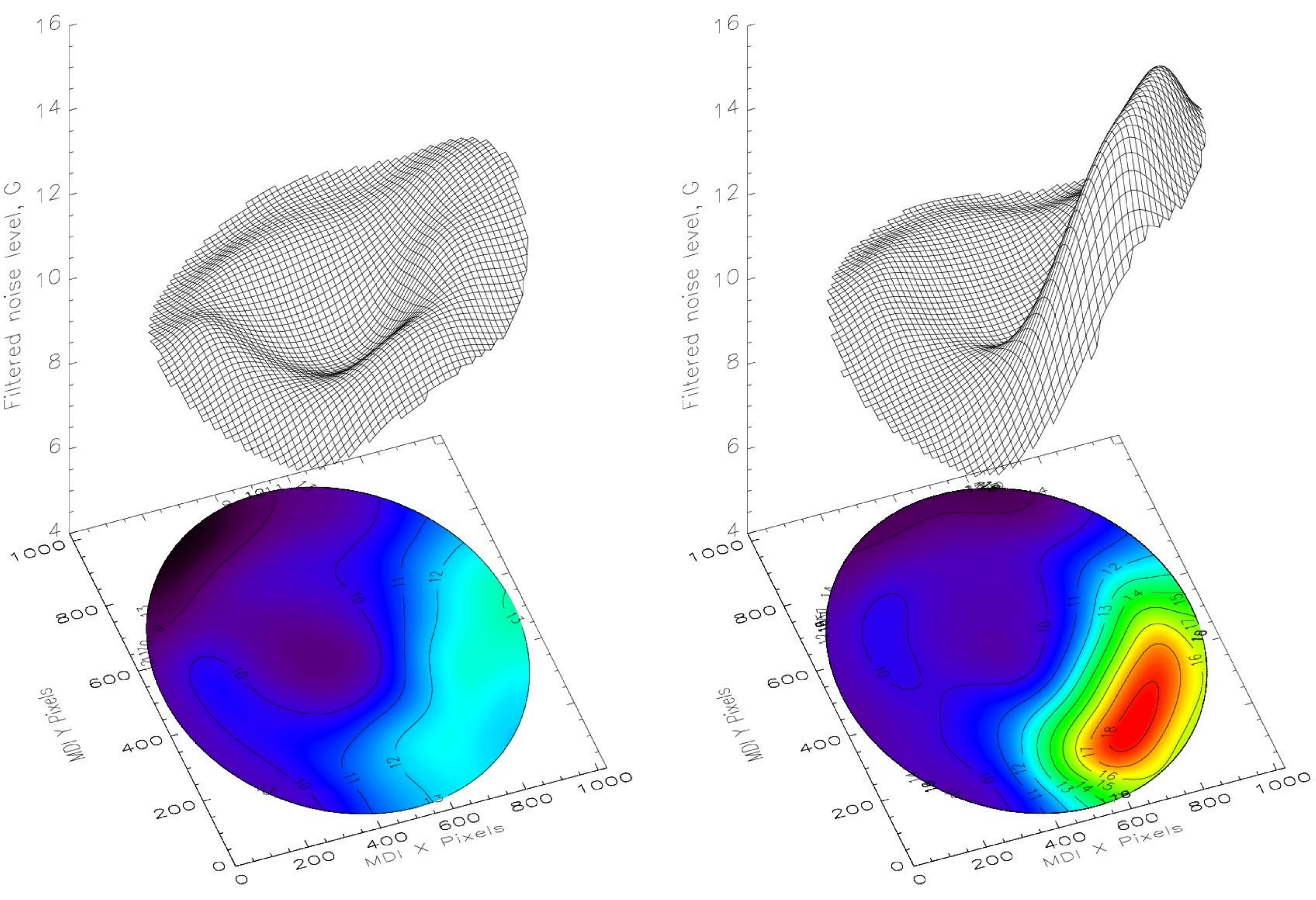}}
 \caption[]{Mean background magnetic flux distribution of magnetograms from 1996-7 (left) and 2008 (right). Both plots have the same colour scale and contours are in 1 G levels.}
 \label{fig:lowfluxdist}
\end{figure*}

It is not immediately clear which approach, scaling or shifting, should be adopted. Therefore, we use the TSI composites as a guide, assuming that when they agree they are a reasonable estimate of the true value of TSI at that time. In Fig.~\ref{fig:shift_multi_plots}b, the ACRIM (yellow) and PMOD (red) composites are also shown. The IRMB composite is no longer extended back to 1978 and so is not considered here. For the period of time covered in this plot, ACRIM and PMOD use the same TSI data sources post-1983, though the way they have been combined is different and after 1989, the trends clearly differ. Therefore, the agreement over 1984 - 1989 is expected. Even though ACRIM and PMOD use different datasets prior to 1984, the two composites agree well between 1981 and 1983. Before 1981, instrument degradation corrections are applied to PMOD, but not to ACRIM. Neither scaling nor shifting fully corrects the KP/512 magnetograms (private communication, Jack Harvey), though together they should give a good indication of the expected range of plausible reconstructions, with shifting minimising the cycle amplitude and scaling maximising the cycle amplitude. As the reconstruction with a shift applied clearly agrees better with ACRIM and PMOD over the 1981 to 1989 period than using the scaling, we choose to adopt the shift for the final reconstruction (see section~\ref{recon:compare}). This, then, assumes that shifting is a more appropriate approach to reconstructing TSI than using a scaling. When considering the range of uncertainty on the reconstruction, we include both the shift and scaling as discussed in section~\ref{recon:uncert}.

\subsection{SoHO/MDI magnetograms}
\label{sec:mdimag}
The two KP datasets, KP/512 and KP/SPM, have been set up to use the same parameters. But to combine MDI with these datasets, an overlap period is needed to find the appropriate B$_{\mathrm{sat}}$ through linear regression. It is, therefore, important that the images used in SATIRE-S are stable over time. For MDI this does not appear to be the case.

The lowest level of magnetic flux in the MDI magnetograms, B$_{\mathrm{thr,MDI}}$ = 3$\sigma_{\mathrm{noise,MDI}}$ is dependent on disk position. The changes to the MDI magnetograms from level 1.5 to 1.8.2 required a re-examination of the mean background flux to recalculate $\sigma_{\mathrm{noise,MDI}}$ as a function of disk-position.

We used a similar approach to \cite{OrtizSolanki2002} and applied this to two sets of 29 magnetograms from both 1996/7 and 2008, chosen for the minimal activity present on the disk. The procedure is as follows. A 100 $\times$ 100 pixel box centred on every pixel in the magnetogram was used to determine if the value of the central pixel exceeded three standard deviations of the mean. If the pixel did exceed this threshold, it was set to the mean. This process was repeated two further times and then the median of the 29 distributions was taken. The contour and surface plots of the level 1.8.2 magnetograms in Fig.~\ref{fig:lowfluxdist} show the mean background flux distribution in 1996-7 (left) and 2008 (right).

Two things are clear from Fig.~\ref{fig:lowfluxdist}: the distribution is not uniform across the disk and it changed between 1996 and 2008. The non-uniformity across the disk, seen as an increase from the upper-left to lower-right in both plots, can possibly be explained by the inhomogenous nature of the Michelson interferometers in the field of view: slightly different wavelengths are observed across the disk. However, the increase over time of the lower-right quadrant is of most concern here. Let us now consider possible causes for such a change.

\begin{figure*}[t]
 \resizebox{\hsize}{!}{\includegraphics[bb=40 25 600 300,angle=0]{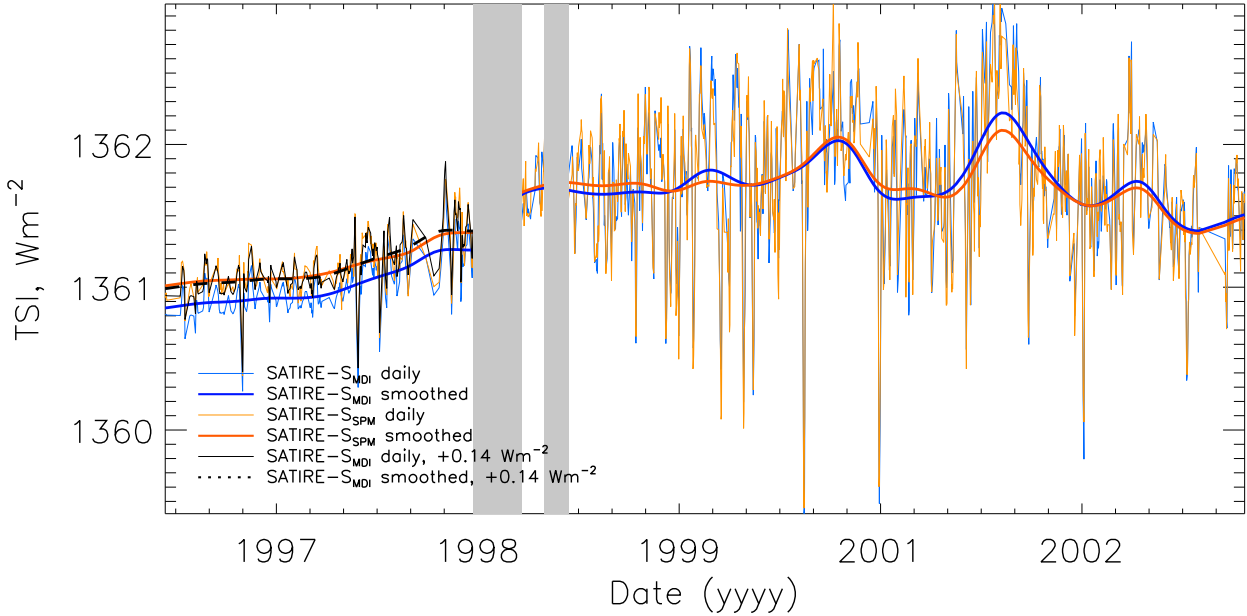}}
 \caption[]{Reconstructions of KP/SPM (orange) and SoHO/MDI (blue) for the overlap period May 1996 to September 2003; KP/SPM is normalised to MDI after February 1999. Daily data are shown as thin lines, smoothed as thick, with the grey boxes covering periods when SoHO was out of contact. Prior to the loss of contact, MDI is also plotted shifted by a constant value (daily data, black thin solid line; smoothed, dashed thick line).}
 \label{fig:shift_pre_holiday}
\end{figure*}

Aside from several shorter periods without observation, there were two extended periods, between June and October 1998 and from December 1998 to February 1999, when SoHO lost contact and was in an uncontrolled state. During the early years of operation, prior to this loss of contact, the focus setting of MDI was deliberately set to ensure the quality of the `high resolution field' mode observations and, as a result, `full-disk' mode was slightly defocused. When contact was re-established, following an analysis that showed this defocusing was not required, the instrument was fully refocused, which significantly changes the point spread function (PSF). Subsequent periodic refocusing was performed to correct a drift and maintain the focus, which should basically maintain the PSF at roughly the level achieved after the loss of contact. This, among other effects such as e.g. the front window darkening, thermal effects, a small change in the observed wavelength and any unknown effects during the period without contact, may have contributed to a change in the registered magnetic flux as a function of disk-position (private communication, Rock Bush). We found that dates when focus changes were performed correlate with a change in distribution of flux in the lower right quadrant. Indeed, by bringing the instrument into focus and narrowing the PSF, the expected outcome would be an increase in the observed magnetic flux, and this is exactly what can be seen in Fig.~\ref{fig:lowfluxdist}. The changes warrant a deeper investigation of the magnetogram changes (see forthcoming thesis by Will Ball). Here, we are mainly interested in global sensitivity to the magnetic flux through the effect they have on a reconstructed irradiance, which can be judged by comparing reconstructions using MDI and KP/SPM data prior to and after the loss of telemetry.

In Fig.~\ref{fig:shift_pre_holiday}, daily and smoothed reconstructions for the overlap period of MDI (blue) and KP/SPM (orange) are shown. The two reconstructions have a unity regression (see section~\ref{recon:combine}) post-SoHO gaps (grey shading), but show a shift prior to the gaps. It can be rectified by a simple shift in the MDI-based recnstruction of 0.14 Wm$^{-2}$. The shifted reconstruction is shown as daily (thin, black) and smoothed (dashed) lines. The agreement of the trends in the shifted MDI and KP/SPM reconstructions is excellent.

Note that a defocus also reduces contrast in the continuum intensity images \citep{DanilovicGandorfer2008}, which would also introduce a bias between sunspot and umbral areas if determined from the period prior to loss of contact relative to determining it afterward. We chose thresholds based on continuum intensity images from after the loss of telemetry (see section~\ref{homog:spot}), so this is not a concern here.

With both of these results, along with the knowledge that the source of the difference almost certainly lies with MDI given the known changes over the SoHO data gaps, we decide to only use MDI data from after the data gaps and use KP/SPM to reconstruct prior to this.

\section{Reconstruction of TSI for cycles 21-23}
\label{sec:recon}
In this section, we explain how we produce the three-cycle reconstruction, independently of any TSI composite after 1989, and how the errors have been formulated. The ultimate aim here is to produce a composite-independent reconstruction of TSI. The SORCE/TIM data, unused in the composites at the time of writing, is therefore an ideal candidate for use as an independent basis to fix the model's free parameter.

\subsection{Combining MDI and KP/SPM data}
\label{recon:combine}
\begin{figure*}[!ht]
 \resizebox{\hsize}{!}{\includegraphics[bb=80 280 725 550, angle=180]{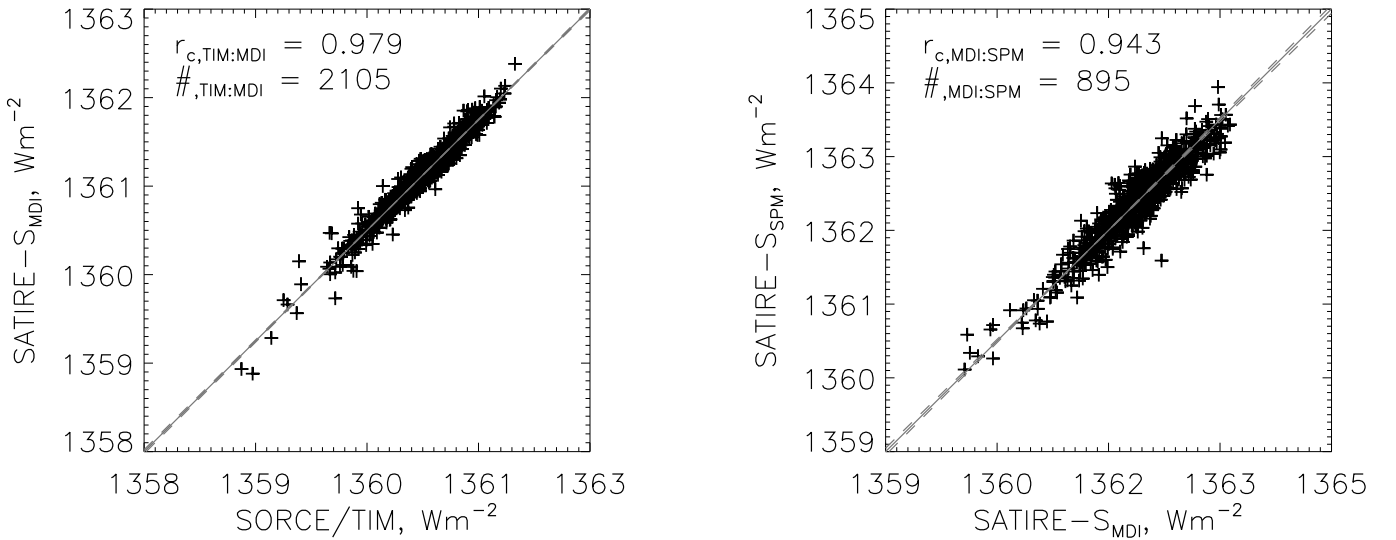}}
 \caption[]{Regression of (left) TIM and MDI and (right) MDI and KP/SPM. The regression line is shown by the solid grey line and 1$\sigma$ uncertainty is represented by the dashed grey lines.}
 \label{fig:tim_mdi_reg}
\end{figure*}

SATIRE-S$_{\mathrm{MDI}}$ is normalised to TIM for 2105 days between 2003 February 25 and 2009 October 31. B$_{\mathrm{sat,MDI}}$ is varied until the best fit regression with a slope of 1.00 is found at 443 G, using the FITEXY linear regression routine \citep{NumRecC}. This is shown as the solid line in the left regression plot of Fig.~\ref{fig:tim_mdi_reg}. The correlation coefficient at this value of  B$_{\mathrm{sat,MDI}}$ is r$_{\mathrm{c,TIM:MDI}}$ = 0.979.

Note that a 30-day period centred on the huge sunspot group passage that reached disk centre around 2003 October 29 has been omitted from this fit. We find that the fit is relatively stable with B$_{\mathrm{sat,MDI}}$ = 443 G for the entire overlap period with TIM except during this passage. SATIRE-S generally reproduces the influence of sunspots well, but this sunspot group seems to have been so exceptionally large that the assumed sunspot properties appear to have changed. SATIRE-S significantly overestimates the darkening produced by this sunspot group and this leads to a change in B$_{\mathrm{sat,MDI}}$ that causes the rest of the time series to be less well reproduced. The spot is only removed when determining B$_{\mathrm{sat,MDI}}$, but considered in the comparisons with the TSI composites (section~\ref{recon:compare}). 

Next, we combine SATIRE-S$_{\mathrm{SPM}}$ with SATIRE-S$_{\mathrm{MDI}}$ using the overlap period of 895 days between 1999 February 18 and 2003 September 21. SATIRE-S$_{\mathrm{SPM}}$ is normalised to SATIRE-S$_{\mathrm{MDI}}$. Using SATIRE-S$_{\mathrm{MDI}}$ with B$_{\mathrm{sat,MDI}}$ = 443 G, a unity regression fit for SATIRE-S$_{\mathrm{SPM}}$ is achieved when B$_{\mathrm{sat,SPM}}$ = 330 G. The regression plot is shown on the right-hand side of Fig.~\ref{fig:tim_mdi_reg}. The correlation coefficient of this fit, r$_{\mathrm{c,MDI:SPM}}$, is 0.943.

With SATIRE-S$_{\mathrm{SPM}}$ established, the link to SATIRE-S$_{\mathrm{512}}$ is automatically prescribed. The same normalisation factor found to bring SATIRE-S$_{\mathrm{SPM}}$ in to line with SATIRE-S$_{\mathrm{MDI}}$ is used for SATIRE-S$_{\mathrm{512}}$. The conversion factor, f$_{\mathrm{512}}$, for B$_{\mathrm{sat,512}}$ = 330 G (see Fig.~\ref{fig:histeq2}b) is 1.580, to the nearest 0.005. The entire time series is normalised to TIM at the December 2008 minimum as a reference period.

Note that the method to combine MDI with KP/SPM differs to that of combining KP/SPM with KP/512. In the latter case, magnetograms from both datasets are compared and adjusted so that the same value of the free parameter can be used. This method is not possible with MDI because the spatial response is not uniform, as discussed in section~\ref{sec:mdimag}.

We use the TSI composites to justify the addition of a constant to pre-1990 KP/512 magnetograms (see section~\ref{homog:shiftcorr}) and, therefore, the reconstruction is independent of the composites after 1989. We denote this full reconstruction as SATIRE-S$_{\mathrm{Ind}}$. In Fig.~\ref{fig:fullrecon} the complete SATIRE-S$_{\mathrm{Ind}}$ reconstruction is shown as daily (thin lines) and smoothed (grey lines) time series. All that remains is to estimate an uncertainty on this result.

\subsection{Uncertainty estimation on the reconstruction}
\label{recon:uncert}

While errors associated with the continuum intensity images and magnetograms enter into the SATIRE-S reconstruction, the largest uncertainties arise in the calculation of the emergent intensities for the different components, in particular for the faculae. These uncertainties are difficult to quantify, though it turns out that changes in the facular contrast are largely absorbed by changes in B$_{\mathrm{sat}}$ as far as TSI reconstructions are concerned.

We therefore try and estimate the uncertainty in the SATIRE-S reconstructions by exploring the change in the modelled TSI that results from variations in the free parameter, B$_{\mathrm{sat}}$, that fall within the allowed fit parameters, such as e.g., quantified by a 1$\sigma$ change in the regressions. We use the difference between the datasets being regressed as an estimate of the standard deviation for both datasets in the FITEXY routine.

Starting with SATIRE-S$_{\mathrm{MDI}}$ we found that, the formal 1$\sigma$ error on the unity fit gradient with SORCE/TIM, $\pm$0.006, corresponds to $\pm$5 G in B$_{\mathrm{sat,MDI}}$. This uncertainty also encompases the best value for B$_{\mathrm{sat,MDI}}$ when a least-squares fit, rather than a regression, is used to find agreement between TIM and SATIRE-S.

The uncertainty over the KP/SPM part of SATIRE-S$_{\mathrm{Ind}}$ is determined from the uncertainty range on the unity regression fit between SATIRE-S$_{\mathrm{MDI}}$ and TIM. In other words, SATIRE-S$_{\mathrm{SPM}}$ is fitted to SATIRE-S$_{\mathrm{MDI}}$ using B$_{\mathrm{sat,MDI}}$ = 438 and 448 G for the upper and lower boundaries, respectively. Then, the 1$\sigma$ errors on these unity regression fits are used to find an expanded uncertainty range prior to 1999. The 1$\sigma$ error on the unity regression gradient is 0.017 and the resulting B$_{\mathrm{sat,SPM}}$ is 307 G and 365 G for the upper and lower range, respectively.

The uncertainty range of B$_{\mathrm{sat,512}}$ in SATIRE-S$_{\mathrm{512}}$ includes the standard error of approximately $\pm$0.04 on the associated conversion factor, f$_{\mathrm{512}}$ (see section~\ref{homog:f512}). We take values of f$_{\mathrm{512}}$ that exacerbate the uncertainty range. For B$_{\mathrm{sat,512}}$ = 365 G, this equates to f$_{\mathrm{512}}$ = 1.600 and for B$_{\mathrm{sat,512}}$ = 307 G, f$_{\mathrm{512}}$ = 1.550.

Finally, we must consider the effect of using a scaling or constant shift correction to KP/512 magnetograms prior to 1990. We consider both approaches so that on each date the pre-1990 correction that produces the largest deviation from SATIRE-S$_{\mathrm{Ind}}$, either highest or lowest in value, is adopted as the upper and lower uncertainty, respectively, to maximise the range.

\begin{figure*}[!ht]
 \resizebox{\hsize}{!}{\includegraphics[bb=50 275 725 560, angle=180]{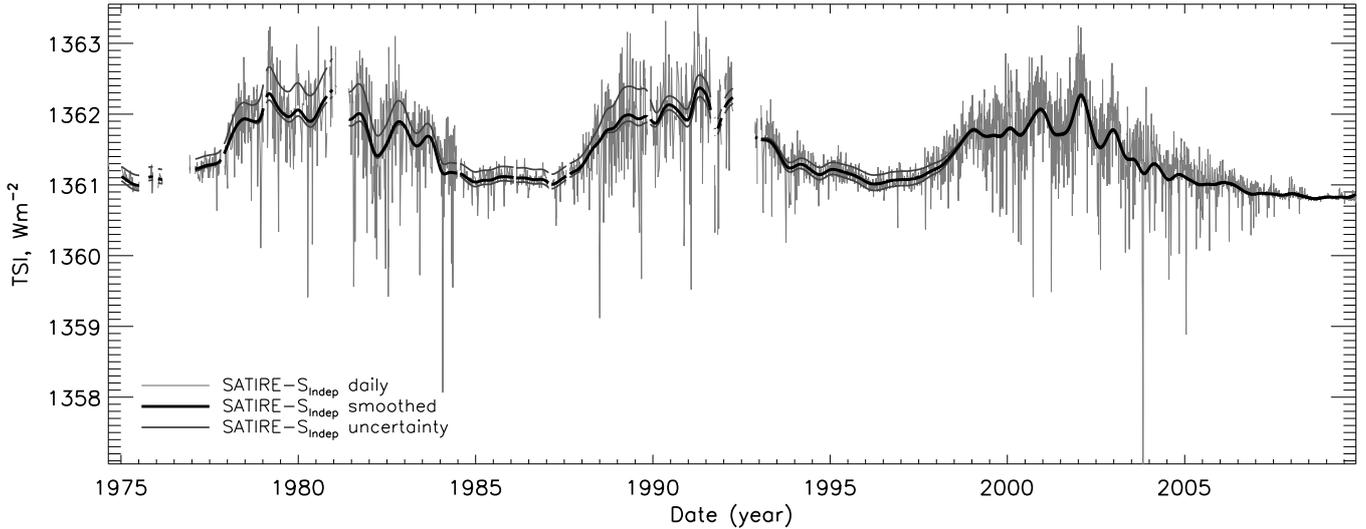}}
 \caption[]{The entire SATIRE-S$_{\mathrm{Ind}}$ dataset. Daily data are shown as solid grey lines when dates are contiguous. The thick black line is smoothed data and the uncertainty range is shown, smoothed only, as thin black lines. Gaps in the curves are present when data gaps are larger than 27 days.}
 \label{fig:fullrecon}
\end{figure*}

The solar minimum of December 2008 is used as the reference date, assuming no long-term uncertainty at that time. Therefore, the two extreme reconstructions that constitute the uncertainty range are normalised to that time, though this does not mean that there is no uncertainty on the absolute value of TSI then. The resulting reconstruction is shown in Fig.\ref{fig:fullrecon} along with the smoothed uncertainty range shown as a contour around the smoothed dataset. While SATIRE-S$_{\mathrm{Ind}}$ pre-1990 is not independent of the TSI composites, the uncertainty range is as the uncertainty includes both the maximum range produced by shifting or scaling KP/512 magnetograms prior to 1990.

\section{Comparison of SATIRE-S$_{\mathrm{Ind}}$ with TSI Composites}
\label{recon:compare}
In this section we compare SATIRE-S$_{\mathrm{Ind}}$ to the three TSI composites in detail. This returns us to a debate over which composite (ACRIM, IRMB or PMOD) provides the best record of TSI. \cite{WenzlerSolanki2009} used SATIRE-S to do this by optimising the model for each composite and found that SATIRE-S fitted to PMOD gave the best correlation coefficients, while ACRIM and IRMB displayed trends that PMOD and SATIRE-S did not show. 

\begin{table*}[t]
\caption[]{Comparison of SATIRE-S$_{\mathrm{Ind}}$ with TSI composites using the original time series (upper half of the table) and detrended (lower half). The table is further split into two to allow a fair comparison with the shorter IRMB composite. Comparisons are made only for dates when data in all datasets are available.}
\begin{center}
\begin{tabular}{llccc}
\hline
\multicolumn{5}{c}{Daily dataset comparison (1978 - 2009)} \\
\hline
Series 1 &   Series 2  &  $r_{c}   [r_{c}^{2}]$  &  Slope  & No. points \\
\hline
ACRIM & SATIRE-S$_{\mathrm{Ind}}$ & 0.879 [0.772] & 0.832 & 5663 \\
PMOD  & SATIRE-S$_{\mathrm{Ind}}$ & 0.936 [0.875] & 1.000 & 5663 \\
ACRIM & PMOD & 0.866 [0.750] & 0.830 & 5663 \\
\hline 
\multicolumn{5}{c}{Daily dataset comparison (1984 - 2009)} \\
\hline 
ACRIM & SATIRE-S$_{\mathrm{Ind}}$ & 0.903 [0.815] & 0.835 & 5150 \\
IRMB  & SATIRE-S$_{\mathrm{Ind}}$ & 0.924 [0.854] & 0.978 & 5150 \\
PMOD  & SATIRE-S$_{\mathrm{Ind}}$ & 0.942 [0.888] & 0.992 & 5150 \\
ACRIM & IRMB & 0.924 [0.853] & 0.856 & 5150 \\
PMOD  & IRMB & 0.922 [0.850] & 1.013 & 5150 \\
ACRIM & PMOD & 0.894 [0.799] & 0.841 & 5150 \\
\hline 
\hline
\multicolumn{5}{c}{Daily dataset comparison, detrended (1978 - 2009)} \\
\hline
ACRIM & SATIRE-S$_{\mathrm{Ind}}$ & 0.866 [0.750] & 0.986 & 5663 \\
PMOD  & SATIRE-S$_{\mathrm{Ind}}$ & 0.873 [0.762] & 1.039 & 5663 \\
ACRIM & PMOD & 0.830 [0.689] & 0.947 & 5663 \\
\hline 
\multicolumn{5}{c}{Daily dataset comparison, detrended (1984 - 2009)} \\
\hline 
ACRIM & SATIRE-S$_{\mathrm{Ind}}$ & 0.889 [0.791] & 0.992 & 5150 \\
IRMB  & SATIRE-S$_{\mathrm{Ind}}$ & 0.940 [0.883] & 1.036 & 5150 \\
PMOD  & SATIRE-S$_{\mathrm{Ind}}$ & 0.877 [0.769] & 1.032 & 5150 \\
ACRIM & IRMB & 0.945 [0.892] & 0.957 & 5150 \\
PMOD  & IRMB & 0.911 [0.831] & 0.994 & 5150 \\
ACRIM & PMOD & 0.847 [0.718] & 0.960 & 5150 \\
\hline \end{tabular}
\end{center}
\label{tab:comp_compare}
\end{table*}

\begin{figure*}[!ht]
 \resizebox{\hsize}{!}{\includegraphics[bb=50 340 750 550, angle=180]{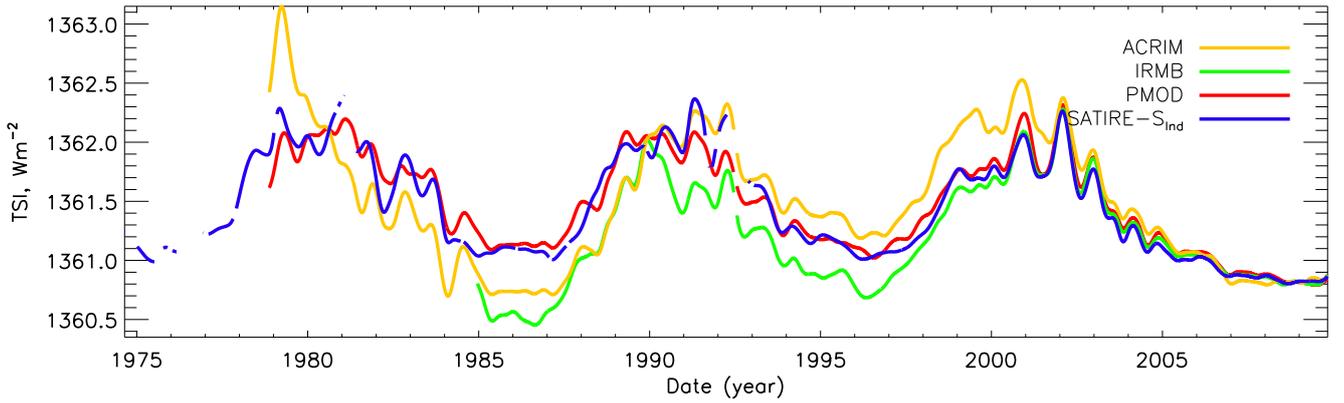}}
 \caption[]{Comparison of smoothed TSI composites and SATIRE-S$_{\mathrm{Ind}}$, all normalised to SORCE/TIM at the minimum of December 2008. ACRIM is in yellow, IRMB in green, PMOD in red and SATIRE-S$_{\mathrm{Ind}}$ in blue. Gaps in the curves are present when data gaps are larger than 27 days.}
 \label{fig:composites}
\end{figure*}

\begin{figure*}[!ht]
 \resizebox{\hsize}{!}{\includegraphics[bb=50 300 775 585, angle=180]{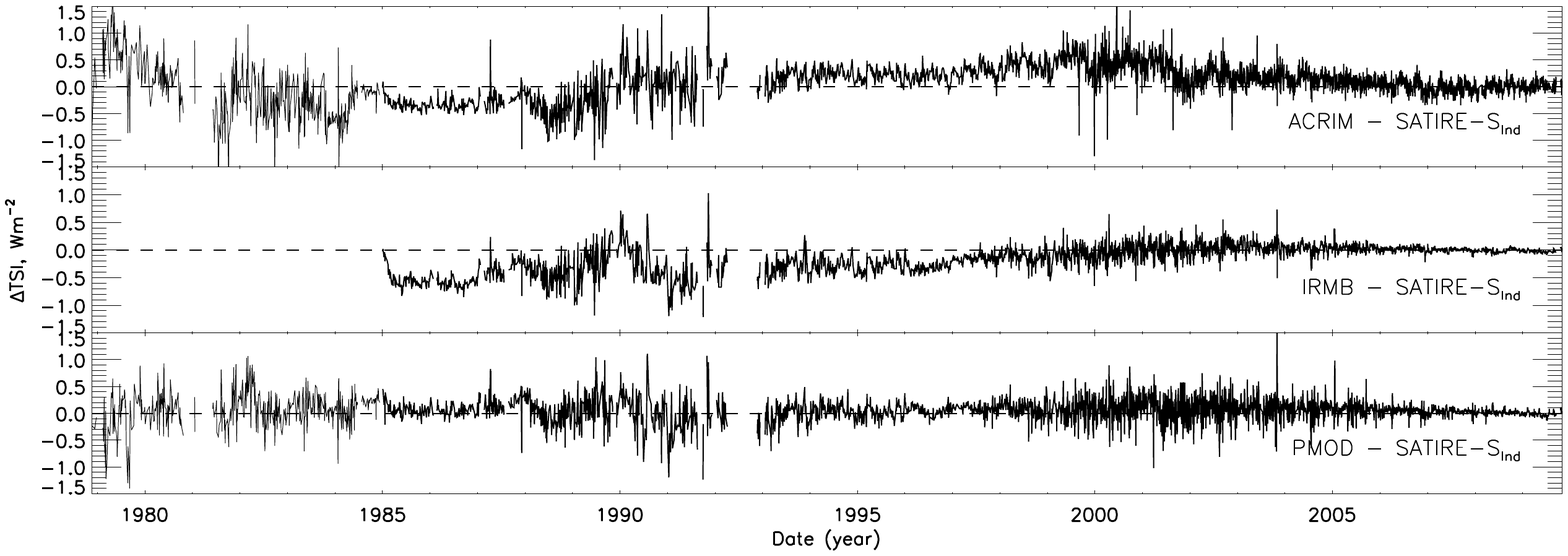}}
 \caption[]{Difference plots of (\textbf{top}) ACRIM - SATIRE-S$_{\mathrm{Ind}}$, (\textbf{middle}) IRMB - SATIRE-S$_{\mathrm{Ind}}$, (\textbf{bottom}) PMOD - SATIRE-S$_{\mathrm{Ind}}$. The plots cover the period 1984-2009 for dates when data are available in all four datasets and prior to this when data are available in all datasets excluding IRMB.}
 \label{fig:deltatsi}
\end{figure*}

SATIRE-S$_{\mathrm{Ind}}$ provides a much stronger test as it covers a longer period and is independently optimised. In Fig.~\ref{fig:composites}, all data are shown together, normalised to the mean value of SORCE/TIM (not shown) over the six month period centred on the December 2008 solar minimum. The data have been smoothed to aid clarity, with PMOD in red, IRMB in green, ACRIM in yellow and SATIRE-S$_{\mathrm{Ind}}$ in blue. In Fig.~\ref{fig:deltatsi}, the difference between the composites and SATIRE-S$_{\mathrm{Ind}}$ are shown for daily data for the period 1978-2009. Before 1984, data are shown on each date that data are available in PMOD, ACRIM and SATIRE-S$_{\mathrm{Ind}}$; after 1984, data are shown when available in all four datasets. Table~\ref{tab:comp_compare} summarises comparisons between the TSI composites and SATIRE-S$_{\mathrm{Ind}}$. It is divided in two halves deliniated by the double line: the top half shows the results of using the daily dataset while the bottom half is for detrended data only so that the rotational variability alone can be considered. Detrending in all cases has been done by subtracting the smoothed dataset from the original. These two halves are further divided into comparisons spanning 1978-2009 and 1984-2009, the latter period being the one to which the IRMB composite is restricted. All comparisons use only dates when data are available in all datasets. This means that the correlation coefficients, r$_{\mathrm{c}}$, and slopes listed in Table~\ref{tab:comp_compare} are strongly weighted by cycle 23, as most data points in SATIRE-S$_{\mathrm{Ind}}$ are during that period.

In Fig.~\ref{fig:composites}, PMOD displays a smaller trend than the other two composites at nearly all times. Also, SATIRE-S$_{\mathrm{Ind}}$ is in good agreement with PMOD at all times, except around the peak of cycle 22. SATIRE-S$_{\mathrm{Ind}}$ agrees better with PMOD than with other TSI composites and better than the composites agree with each other, reflected in the 1984-2009 daily non-detrended results. Since the end of cycle 21, SATIRE-S$_{\mathrm{Ind}}$ shows a decrease in minima levels within the uncertainty range. This secular change appears to be most significant between the minima of cycle 23, in agreement with PMOD, and does not show the more marked secular variations as in the IRMB and ACRIM composites. Interestingly, IRMB and SATIRE-S$_{\mathrm{Ind}}$ appear to have the best agreement during the declining phase of cycle 23, with a gradient in better agreement with SORCE/TIM than either PMOD or ACRIM (for SATIRE-S this is by construction). Note that the regressions and correlation coefficients improve in the shorter, later period from 1984, consistent with the fact that the older data prior to 1984 are less certain.

Considering the detrended TSI comparisons between 1984 and 2009, which are measures of rotational modulation of irradiance and the effects of active region evolution, agreement between all datasets is generally good. IRMB has the best agreement with SATIRE-S$_{\mathrm{Ind}}$, followed by ACRIM and finally PMOD. The agreement between SATIRE-S$_{\mathrm{Ind}}$ and IRMB is almost as good as the best agreement between TSI composites, which is found between IRMB and ACRIM. It is also clear that the rotational scatter in Fig.~\ref{fig:deltatsi} between IRMB and SATIRE-S$_{\mathrm{Ind}}$ during cycle 23 is much lower than with the other TSI composites though the longer-term trend during the rising phase is not the same, especially before 1997; this entire period utilises the DIARAD instrument on SoHO.

The high level of agreement in the detrended time series between SATIRE-S$_{\mathrm{Ind}}$ and IRMB means that SATIRE-S$_{\mathrm{Ind}}$ is able to recreate rotational variability to the level of uncertainty between the TSI composites. This goes some way to explaining why IRMB has a relatively good regression gradient with SATIRE-S$_{\mathrm{Ind}}$ when considering the non-detrended data, despite the large disagreement in the smoothed long-term trend shown in Fig.~\ref{fig:composites}.

The regression slopes also indicate that the rotational variation in SATIRE-S$_{\mathrm{Ind}}$ falls in between the TSI composites, with ACRIM having the largest amplitude, and PMOD and IRMB the lowest. It should be borne in mind that these results may be affected strongly by active regions. It is well documented \citep{UnruhKrivova2008, BallUnruh2011} that TSI can be overestimated in SATIRE-S when large facular regions transit the edge of the solar disk. Some of this overestimation has been improved upon through the modification described in section~\ref{homog:darkpix}. It may be the case, though, that SATIRE-S$_{\mathrm{Ind}}$ still overestimates rotational varability and that is why the regression slopes suggest that SATIRE-S$_{\mathrm{Ind}}$ has a higher amplitude than PMOD and IRMB.

In general, the composites and reconstruction agree well with each other on rotational time scales. The major cause of differences between composite datasets are likely to be produced by the use of different data sources in cycle 23. Assuming, for the sake of argument, that SATIRE-S$_{\mathrm{Ind}}$ is a good representation of TSI behaviour we find that PMOD is the most reliable composite of TSI, taking into account rotational, cyclical and secular variations, although on rotational time scales IRMB is superior.

\begin{table*}[t]
\caption[]{Detailed comparison between PMOD and SATIRE-S considering only dates when both datasets overlap$^{1}$.}
\begin{center}
\begin{tabular}{lccccccc}
\hline
\multicolumn{8}{c}{Cycle Maximum$^{2}$ and Minimum TSI, Wm$^{-2}$} \\
\hline 
Series & Min C20/21$^{3}$ & Max C21 & Min C21/22 & Max C22 & Min C22/23 & Max C23 & Min C23/24 \\ \hline 
SATIRE-S$_{\mathrm{Ind}}$ & 1361.06 & 1362.01 & 1361.08 & 1361.96 & 1361.01 & 1361.74 & 1360.82 \\
PMOD  & - & 1362.02 & 1361.14 & 1362.10 & 1361.07 & 1361.82 & 1360.82 \\
\hline
\multicolumn{8}{c}{Cycle Maximum and Minimum TSI Relative to 2008 Minimum, Wm$^{-2}$} \\
\hline 
Series & Min C20/21 & Max C21 & Min C21/22 & Max C22 & Min C22/23 & Max C23 & Min C23/24 \\ \hline 
SATIRE-S$_{\mathrm{Ind}}$ & 0.24$^{+0.14}_{-0.06}$ & 1.19$^{+0.40}_{-0.09}$ & 0.26$^{+0.14}_{-0.06}$ & 1.14$^{+0.42}_{-0.09}$ & 0.20$^{+0.12}_{-0.09}$ & 0.92$^{+0.02}_{-0.02}$ & 0.00$^{+0.00}_{-0.00}$ \\
PMOD  & - & 1.21 & 0.32$\pm$0.01 & 1.28 & 0.25$\pm$0.10 & 1.00 & 0.00$\pm$0.16 \\
\hline 
\end{tabular}
\begin{tabular}{lccccc}
\multicolumn{6}{c}{Statistics of Full and Cycle Periods} \\
\hline 
Period & Start date & End date & $r_{c}   [r_{c}^{2}]$ & Slope & No. points \\ \hline 
Full & 1978/12/10 & 2009/10/31 & 0.958 [0.919] & 0.995 & 5899 \\
Cycle 23 & 1996/05/15 & 2008/12/15 & 0.981 [0.963] & 0.978 & 3760 \\
Cycle 22 & 1986/09/15 & 1996/05/14 & 0.917 [0.840] & 1.040 & 1148 \\
Cycle 21 (part) & 1978/12/09 & 1986/09/14 & 0.898 [0.806] & 1.050 & 740 \\
\hline 
\end{tabular}
\end{center}
\tablefoot{
$^{1}$ The uppermost table shows the absolute values at activity cycle maximum and minimum. The middle table gives irradiance values relative to the December 2008 minimum with uncertainty estimates from SATIRE-S$_{\mathrm{Ind}}$ and errors from PMOD. The bottom set give the regression slopes and correlation coefficients from the different cycles of the reconstruction. $^{2}$ TSI values are calculated from $\pm$90 days centred on the dates of maximum and minimum smoothed monthly sunspot number from SIDC data. $^{3}$ Only SATIRE-S$_{\mathrm{Ind}}$ data exist around the minimum of cycle 20/21 minimum. Minimum values are estimated between June 1975 and July 1976 due to the lack of data around the sunspot minimum of June 1976.}
\label{tab:max_min_tsi}
\end{table*}

\begin{figure*}[!ht]
 \resizebox{\hsize}{!}{\includegraphics[bb= 40 40 800 415, angle=180]{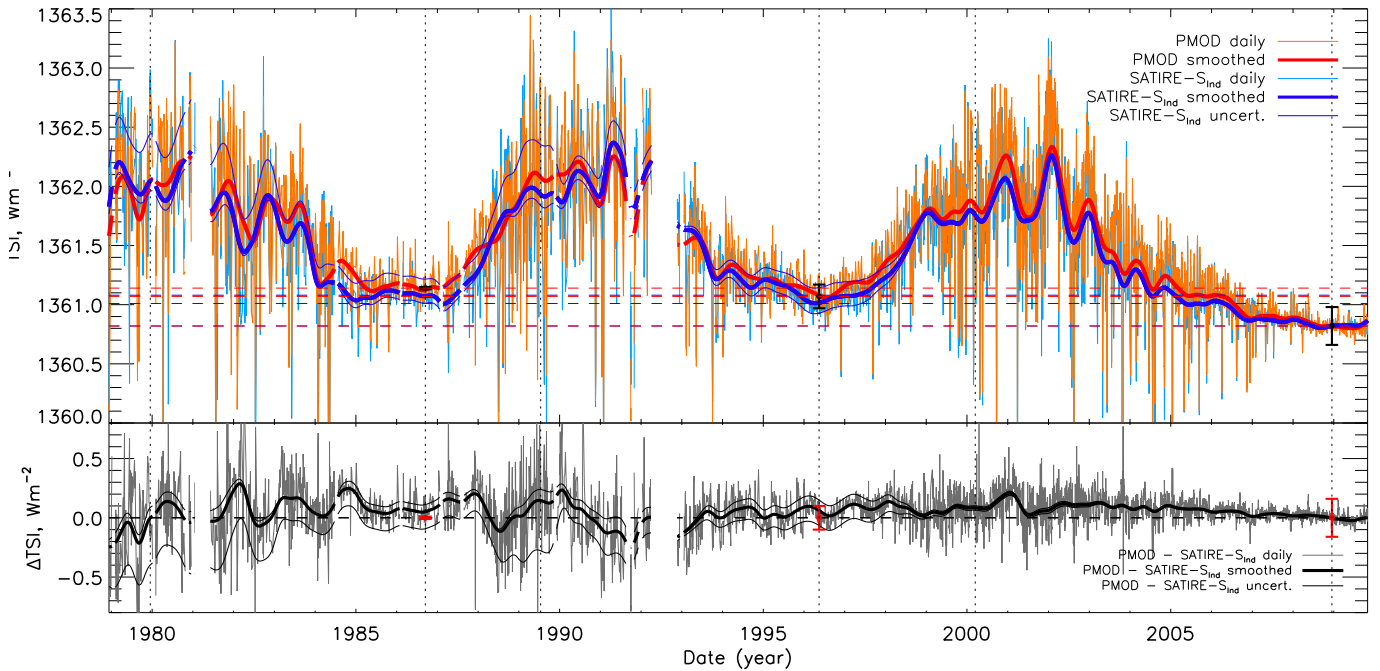}}
 \caption[]{In the upper plot, PMOD (daily data, light red; smoothed, thick red line) and SATIRE-S$_{\mathrm{Ind}}$ (daily data, light blue; smoothed, thick blue) between 1978 and 2009 normalised to SORCE/TIM at December 2008. The thin blue lines mark the uncertainty range of SATIRE-S$_{\mathrm{Ind}}$ (only smoothed values plotted). In the lower plot, the difference between PMOD and SATIRE-S$_{\mathrm{Ind}}$ is shown (daily, grey; smoothed, black) along with the difference of the uncertainty with respect to PMOD. The black and red error bars are the errors from \cite{Frohlich2009b} in the upper and lower plots, respectively. Dotted vertical lines indicate cycle maxima and minima. Dashed horizontal lines signifying cycle minima are plotted to aid the reader. Gaps in the curves are present when data gaps are larger than 27 days.}
 \label{fig:pmod_satire}
\end{figure*}

Given that PMOD is the composite by far the closest to SATIRE-S$_{\mathrm{Ind}}$ overall, we concentrate in the following on an in-depth comparison of the twoS.S.Dhomse@leeds.ac.uk. Figure~\ref{fig:pmod_satire} shows in the top panel PMOD (light red, daily; red, smoothed) and SATIRE-S$_{\mathrm{Ind}}$ (light blue, daily; blue, smoothed) irradiance. The bottom plot shows the difference between the two, PMOD minus SATIRE-S$_{\mathrm{Ind}}$ (black, smoothed). Also indicated are the smoothed uncertainty range for SATIRE-S$_{\mathrm{Ind}}$, as thin, light blue and black lines in top and bottom plots respectively, and PMOD black and red error bars from \cite{Frohlich2009b}. The values of these errors are given in the middle table of Table~\ref{tab:max_min_tsi}. These error bars were derived by \cite{Frohlich2009b} as an estimate of the long-term stability in the instruments and the calculation is described within that publication. The error bar at the 1986 minimum is very small because this minimum was used as the reference date for comparison with other minima. In the plot, PMOD is normalised to the mean of SATIRE-S$_{\mathrm{Ind}}$ at the 2008 minimum, 2008 December 15 $\pm$90 days, to help discuss the change between cycle minima and bring irradiance levels in line with SORCE/TIM. Dashed lines in blue and red guide the eye to minima levels in 1986, 1996 and 2008.

The immediate impression is that across the 5899 days, where data are available in both time series, there is very good agreement on all timescales for the majority of the time period. The long-term trend is very similar over the 31 year period of overlap and the comparison shows that SATIRE-S$_{\mathrm{Ind}}$ can explain almost 92\% of variability in the TSI composite\footnote{Note that the correlation coefficients and regression slopes between PMOD and SATIRE-S$_{\mathrm{Ind}}$ differ in Tables~\ref{tab:comp_compare} and \ref{tab:max_min_tsi} due to a change in the data points considered.}.

In the lower part of Table~\ref{tab:max_min_tsi}, comparisons have been made for the full cycle and each individual cycle. The agreement is highest in cycle 23 during which SATIRE-S$_{\mathrm{Ind}}$ is able to explain more than 96\% of the variation. We note that the difference plot shows that the gradient in TSI during the rising and falling phases of cycle 23 is slightly lower in SATIRE-S$_{\mathrm{Ind}}$, a consequence of using TIM to fit the reconstruction, which shows a gentler decline of TSI with time than PMOD.

A significant result is that SATIRE-S$_{\mathrm{Ind}}$ is able to reproduce the secular, inter-cycle, change in PMOD TSI, mimicking the $\sim$0.2 Wm$^{-2}$ decline between 1996 and 2008. There is a slight difference between minima values in 1996, relative to 2008: within the uncertainty range, SATIRE-S$_{\mathrm{Ind}}$ suggests a decline between cycle 23 minima of between 0.11 and 0.32 Wm$^{-2}$, encompassing the result from PMOD. The best fit from SATIRE-S$_{\mathrm{Ind}}$ also falls within the error bars of PMOD in 1996. SATIRE-S$_{\mathrm{Ind}}$ agrees with PMOD on the secular change over cycle 22 and possibly shows a slight secular increase over cycle 21 as can be seen in Table~\ref{tab:max_min_tsi}, though data are sparse prior to 1977 and there is no change within the uncertainty range. These results, summarised in Table~\ref{tab:max_min_tsi}, calculate the 1976 minimum using the mean of data between June 1975 and July 1976, while all other maxima and minima are for six month periods centred on the maximum and minimum months of the sunspot cycle from the Solar Influences Data Centre (SIDC) smoothed monthly sunspot dataset \citep{sidc}.

Although the long-term agreement between PMOD and SATIRE-S$_{\mathrm{Ind}}$ is generally very good, there are periods of one to two years during the maxima of cycles 21 and 22 where the difference between them reaches $\sim$0.3 Wm$^{-2}$, as can be seen in the lower panel of Fig.~\ref{fig:pmod_satire}. This is likely to be the result of the poorer data quality of KP/512 and the larger uncertainty in PMOD during the first two cycles. The increased difference in 1990/1 occurs in KP/512 data in the period where no additive correction is applied to the magnetograms (see section~\ref{homog:shiftcorr}). There is an implication that a scaling correction per se should not only be applied to pre-1990 magnetograms. Only applying a constant shift as a correction is probably also not correct and a combination of the two is probably more appropriate. A fuller investigation should be performed to understand the time-dependent problems in the KP/512 magnetograms, expanding on the work done by \cite{ArgeHildner2002}.

There is also clearly a different six-monthly trend apparent in the period from late-1996 to 1998 where the gradient in SATIRE-S$_{\mathrm{Ind}}$ is lower than in PMOD. Although MDI data over this period have not been employed for the final reconstruction, for this small stretch both SATIRE-S$_{\mathrm{MDI}}$ and SATIRE-S$_{\mathrm{SPM}}$ show a very similar trend (see Fig.~\ref{fig:shift_pre_holiday}). \cite{KrivovaSolanki2011b} discuss this difference in some detail and also compare other radiometric observations that show significant deviations from PMOD. It is therefore very likely that the degradation corrections applied during this period to VIRGO data may profit from being revisited.

\section{Discussion and conclusions}
\label{sec:discuss}

In this paper we present the first reconstruction using the SATIRE-S model to fully cover solar cycles 21, 22 and 23. The reconstruction is compared with three TSI composites constructed by \cite{WillsonMordvinov2003}, \cite{DewitteCrommelynck2004} and \cite{FrohlichLean1998} and referred to as ACRIM, IRMB and PMOD respectively, the last of which is considered in more detail. The modelled time series was constructed without reference to any of the composites, except prior to 1990. This means that the entirety of cycle 23 is independent of all the TSI composites. This provides an opportunity to test the model's assumption: that changes in irradiance are only caused by the evolution of surface magnetic flux.

The SATIRE-S reconstruction is found to agree best with the PMOD composite on time scales longer than rotational. On shorter time scales the IRMB composite displays a better agreement, though all three composites are similar on this shorter time scale. The model is not just extremely successful at recreating rotational and cyclical variability in PMOD, but also in reproducing the secular trend, particularly apparent in the decline in TSI of $\sim$0.2 Wm$^{-2}$ between the minima of cycle 23. The difference between the model and PMOD minima of cycle 21/22 and cycle 22/23, relative to the 2008 minima, is less than or equal to 0.06 Wm$^{-2}$ and within the uncertainty of the model and of the observations. The fact that the model and the PMOD composite agree very well provides support for both of them.

The model is able to account for 92\% of the variability in PMOD since 1978 and 96\% of the variability over cycle 23. Due to the general independence of the model, these results support the PMOD composite as being the most realistic record of TSI. It also suggests that rotational, cyclical and secular variations in TSI are mainly the result of the evolution of photospheric magnetic fields. This finding is in opposition to \cite{Frohlich2009b,Frohlich2011} and \cite{Steinhilber2010}, who proposed that secular changes in TSI are caused by a global temperature change (see also \cite{TappingBoteler2007} and references therein).

It is clear from Fig.~\ref{fig:pmod_satire} that there is disagreement between the model and PMOD on periods of one to two years, most significantly over the maximum years of cycle 21 and 22 during which KP/512 are employed in the model. KP/512 data are subject to many more problems than KP/SPM and SoHO/MDI. These have been discussed in sections~\ref{data:512},~\ref{homog:f512} and~\ref{homog:shiftcorr}.

Some of the differences are also a result of inadequate degradation corrections in PMOD. During the period prior to 1982 the NIMBUS-7/HF radiometer was the only instrument in operation. The degradation correction applied to data from this instrument was based on the work done on the VIRGO/PMO6V radiometer in conjunction with ACRIM I data \citep{Frohlich2003}. While we accept that this is a reasonable correction, we believe that some uncertainty in PMOD remains during this early period.

\cite{KrivovaSolanki2011b} found that the degradation correction of VIRGO during the initial years of observation over 1996-1998 may not be entirely adequate. Our study adds to that comparison, which also used UARS/ACRIM II and DIARAD and PMO6V radiometers within the SoHO/VIRGO instrument. We find that SATIRE-S reconstructions using either SoHO/MDI or KP/SPM images disagree in an almost identical manner to the trend in PMOD during the first 18 months of the rising phase of cycle 23. We did find, however, that the magnetic flux in SoHO/MDI magnetograms is different before and after the period when SoHO was not in operation in 1998. This can be attributed to instrument focus changes. While they can be used in shorter reconstruction stretches which agree with the reconstruction using KP/SPM (see Fig.~\ref{fig:shift_pre_holiday}), MDI magnetograms should not be used to test secular variation, without adequately correcting for this effect.

While TSI is an important observable of changes in solar irradiance, many recent climate modelling attempts attribute solar spectral irradiance as physically important when trying to understand the impact the Sun has on the Earth's climate. By design, SATIRE-S automatically produces spectra over climate-relevant wavelengths. A following publication will analyse and make these data available.

The SATIRE-S$_{\mathrm{Ind}}$ full reconstruction covers 6379 days between 1974 and 2009. The daily and smoothed data, along with the uncertainty range can be accessed at http://www.mps.mpg.de/projects/sun-climate/data.html.

\section*{ACKNOWLEDGMENTS}
We thank Rock Bush for his detailed comments regarding the SoHO/MDI instrument. We also thank Greg Kopp for his comments regarding the SORCE/TIM data and Steven Dewitte for providing IRMB data. Thanks is given to Jeneen Sommers at the Stanford Solar Center for help preparing SOHO/MDI data for download and resolving queries. SOHO Data supplied courtesy of the SOHO/MDI and SOHO/EIT consortia. SOHO is a project of international cooperation between ESA and NASA. 

This work was supported in part by the NERC SolCli consortium grant and the STFC grant ST/I001972/1. This work has also been partly supported by WCU grant No. R31-10016 funded by the Korean Ministry of Education, Science and Technology and the {Deut\-sche For\-schungs\-ge\-mein\-schaft, DFG} project number SO~711/1-3. We also thank the International Space Science Institute (Bern) for giving us the opportunity to discuss this work with the international team on "Observing and Modeling Earth's Energy Flows".

We make the use of version d41\_62\_1003 of the PMOD dataset from PMOD/WRC, Davos, Switzerland, and we acknowledge unpublished data from the VIRGO Experiment on the cooperative ESA/NASA Mission SoHO. For sunspot maxmimum and minimum dates, we used data from the SIDC-team, World Data Center for the Sunspot Index, Royal Observatory of Belgium, Monthly Report on the International Sunspot Number between 1974 and 2009. The online catalogue of the sunspot index can be found at: http://www.sidc.be/sunspot-data/.

We thank the referee, Greg Kopp, for careful consideration of the paper and his helpful comments.

\bibliography{bibfile}
\bibliographystyle{aa}

\end{document}